\documentclass[preprint,showpacs,preprintnumbers,amsmath,amssymb,nofootinbib]{revtex4}

\usepackage{etex}
\usepackage{graphicx}
\usepackage{dcolumn}
\usepackage{bm}
\usepackage{amssymb,amsthm,amscd,amsbsy,array}\usepackage{amsmath,dsfont}

\usepackage{graphics,xcolor}

\usepackage{color}
\newcommand{\red}{\textcolor{red}}  
\newcommand{\blue}{\textcolor{blue}}

\newcommand{\PT}{{P\"oschl - Teller\;}}

\newcommand{\DM}{{Displacement Memory\;}}

\newcommand{\half}{{\scriptstyle{\frac{1}{2}}}}
\def\2{{\half}}

\newcommand{\const}{\mathop{\rm const}\nolimits}

\def\p{{\partial}}

\def\beqa{\begin{eqnarray}}
\def\eeqa{\end{eqnarray}}

\def\barray{\left(\begin{array}}
\def\earray{\end{array}\right)}
\def\barraynb{\begin{array}}
\def\earraynb{\end{array}}

\def\smallover#1/#2{\hbox{$\textstyle\frac{#1}{#2}$}} %

\newcommand{\cA}{{\mathcal{A}}}

\def\smallcirc{{\raise 0.5pt \hbox{$\scriptstyle\circ$}}}
\def\aand{{\quad\text{\small and}\quad}}

\def\ie{{\;\text{\small i.e.}\;}}
\def\ie,{{\;\text{\small i.e.,}\;}}

\def\GWs{{gravitational waves\,}}
\def\VM{{Velocity Effect\,}}
\def\DM{{Displacement Effect\,}}

\def\benu{\begin{enumerate}}
\def\eenu{\end{enumerate}}
\def\bitem{\begin{itemize}}
\def\eitem{\end{itemize}}

\def\besub{\begin{subequations}}
\def\esub{\end{subequations}}

\def\?{{\,\gb{\fbox{\texttt{??}}\;}}\,}

\def\StL{{Sturm-Liouville\,}}
\def\cI{{{\cal I}}}

\usepackage[colorlinks=true, pdfstartview=FitV, linkcolor=blue, citecolor=blue, urlcolor=blue]{hyperref} 

\newcommand{\gb}{\quad\colorbox{green}}

\newenvironment{redtext}{\color{red}}
{\ignorespacesafterend}
\newenvironment{bluetext}{\color{blue}}{\ignorespacesafterend}
\newenvironment{greentext}{\color{green}}{\ignorespacesafterend}
\newenvironment{magentatext}{\color{magenta}}{\ignorespacesafterend}
\newenvironment{cyantext}{\color{cyan}}{\ignorespacesafterend}
\newenvironment{orangetext}{\color{orange}}
{\ignorespacesafterend}

\newcommand{\bmagenta}{\begin{magentatext}}
\newcommand{\emagenta}{\end{magentatext}}
\newcommand{\bcyan}{\begin{cyantext}}
\newcommand{\ecyan}{\end{cyantext}}
\newcommand{\bblue}{\begin{bluetext}}
\newcommand{\eblue}{\end{bluetext}}
\newcommand{\bred}{\begin{redtext}}
\newcommand{\ered}{\end{redtext}}

\newcommand{\bgreen}{\begin{greentext}}
\newcommand{\egreen}{\end{greentext}}
\newcommand{\borange}{\begin{orangetext}}
\newcommand{\eorange}{\end{orangetext}}

\numberwithin{equation}{section}

\let\ssection=\section
\renewcommand{\section}{\setcounter{equation}{0}\ssection}
\newcommand{\beq}{\begin{equation}}
\newcommand{\eeq}{\end{equation}}
\newcommand{\bec}{\begin{center}}
\newcommand{\ec}{\end{center}}

\newcommand{\bigbox}[1]{\fbox{%
\rule[-20pt]{0pt}{45pt}$\;\;\displaystyle{#1}\;\;$}
}

\usepackage{amsthm}

{Corollary}[section]

\begin{document}

\preprint{arXiv: v1}

\title{Memory Effect for deformed gravitational waves}

\author{
P.-M. Zhang$^{1,2}$\footnote{Corresponding author mailto:zhangpm5@mail.sysu.edu.cn},
M. Elbistan$^{2}$\footnote{mailto:mahmut.elbistan@bilgi.edu.tr}
and
P. A. Horv\'athy$^{2,3}$\footnote{mailto:horvathy@univ-tours.fr}
}

\affiliation{
$^1$ School of Physics and Astronomy, Sun Yat-sen University, Zhuhai 519082, (China)
\\
${}^2$ Department of Energy Systems Engineering, Istanbul Bilgi University, 34060, Eyupsultan, Istanbul, (Turkey)
\\
${}^{3}$ 
Universit\'e de Tours, Universit\'e d'Orl\'eans, 
 CNRS, IDP, UMR 7013
  37200 Tours, France
}
\date{\today}
\begin{abstract} 
Deformations of sandwich gravitational waves are considered. Squeezing a wave with pure displacement  (DM) parameters to an impulsive wave, the DM property is lost, leaving us with the velocity effect (VM). For each fixed value of the deformation parameter $p$, DM can be restored by further increasing the amplitude which however diverge when $p\to\infty$, and no impulsive wave is obtained. 
\end{abstract}


\maketitle

\tableofcontents

\section{Introduction: }\label{Intro}

The Memory Effect, proposed to test if Einstein \GWs (GW) do indeed exist \cite{BraTho,Favata}, has attracted renewed attention. 
 
First studies \cite{Ehlers} concerned  the \emph{velocity effect} (VM): particles hit by a sandwich wave fly apart with \emph{non-zero constant} velocity. In their seminal paper
Zel'dovich and Polnarev \cite{ZelPol} suggested instead  \GWs generated by 
 flyby would simply displace the particles, yielding the \emph{Displacement effect} (DM).
DM has been confirmed so far for toy profiles which include a Gaussian, the  \PT potential, their derivatives, the Scarf potential, etc. 
 DM requires that the amplitude take some ``magic value'', for which the Wavezone contains an integer number of half-waves  \cite{DM-1,Jibril,DM-2,Sila-PLB,Approxi}. DM, if confirmed experimentally, would be particularly important as it avoids dissipation by implying  cumulative energy conservation \cite{Benin,DM-1,Maluf,Carneiro}.  
  
\emph{Impulsive waves} \cite{PodolskyVeselyCz98,Pod98,Podolsky:1998sa,
Steinbauer97,Steinbauer98,P-SSS17, ZDHimpulsive,SteinbauerComm} are
 obtained by squeezing smooth profiles to a Dirac delta as it will be recalled in sec.\ref{squeezedSec}. Remarkable observations of Podolsky  \cite{Pod98,PodolskyVeselyCz98}, and of Steinbauer \cite{Steinbauer97},
say, moreover, that the impulsive limit is independent
of  the initial profile, and support only VM trajectories.
In sect. \ref{DMSec} we argue that DM \emph{can} be obtained --- however at the expense of increasing the amplitude 
which then diverges when the deformation parameter goes to infinity~:  no impulsive wave is obtained~\footnote{With the words of late Lochlainn  O'Raifeartaigh~: \emph{you can not eat the cake, and have the cake}.}. Further details will be presented in \cite{ZEH-PR}.
  
\section{Memory in the Eisenhart-Duval framework}\label{BargSec}

Further insight is gained from the ``Kaluza-Klein-type" approach of Eisenhart, and of Duval et al called the (E-D) framework \cite{Eisenhart,DBKP,DGH91}~: 
 the $D+1$ dimensional NR dynamics can be studied by projecting lightlike geodesics in $D+2$ dimensional ``Bargmann'' manifold ${\cal M}=\big\{X,U,V\big\}$ endowed with a metric $g_{\mu\nu}$ for which $\xi=\p_V$ is a covariantly constant null vector. 
 
Henceforth we restrict  our attention at a particle in $D=1$ dimensional flat space moving in a harmonic potential with possibly time-dependent profile $\cA(U)$. The metric is written, in Brinkmann coordinates \cite{EZHRev}, 
\beq
dX^2+2dUdV - \cA(U)\,X^2
\,dU^2\,,
\label{Bplanewave}
\eeq
where $X$ is the space-like transverse coordinate, and $U$ and $V$ are light-cone coordinates  \cite{DM-1,DM-2,GlobalCarroll}. $\cA(U)$ is  the profile of the wave. For initial conditions 
\beq
X(-\infty)=X_0 \aand \dfrac{dX}{dU}(-\infty)=0
\label{initcond}
\eeq 
the geodesics $X(U)=P(U) X_0\,$ are found by solving the \StL eqn~\footnote{Eqn. \eqref{SLeq} can indeed be viewed as a zero-energy Schr\"odinger equation \cite{DM-1}. DM solutions correspond to trajectories with large-$U$ behavior $X(\infty)=\const.$  In quantum language, they are (non-normalizable) bound states \cite{DM-1}.
} 

\beq
\dfrac{d^2\! P}{dU^2} + \cA(U) P = 0\,, 
\qquad
P(-\infty)=1   \aand \frac{dP}{dU}(-\infty)=0\,.
\label{SLeq}
\eeq 

\begin{figure}[h]
\includegraphics[scale=.75]{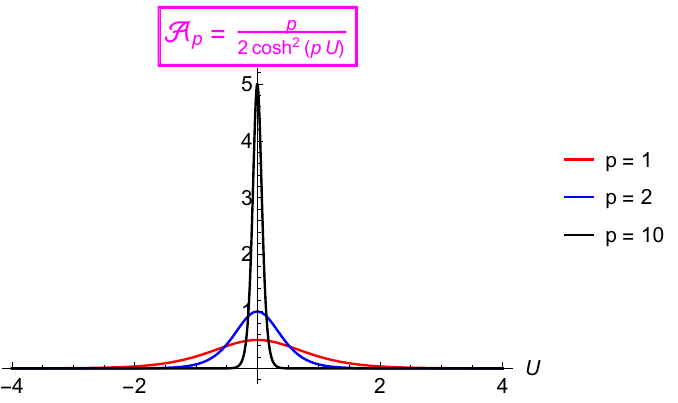}\vskip-5mm\caption{\textit{\small 
For fixed  amplitude $k$, the squeezed \PT profile \eqref{PTp} approaches, for $p\to \infty$, that of an impulsive wave. 
}
\label{squeezePT} }
\end{figure}

 In \cite{Approxi} we considered the {``expanded''}  profile  which generalises  the familiar Gaussian bell ($q=1$), and letting $q\to\infty$ it goes over into a square potential \cite{Kar3,Approxi}. 
In this paper we study instead the behaviour under another, oppositely-oriented deformations of the profile. In detail, we start with a \PT\! profile \cite{PTeller,Khare}
\begin{equation}
 \cA^{PT}(U) = \dfrac{k^{PT}}{2\cosh^2 U}\,,
\label{PTGprof}
\end{equation}
where  the $k^{PT}\equiv k_{m}=2m(m+1)$ with $m$ an integer are \emph{constant} ``magic'' DM  amplitudes \cite{DM-1,DM-2}.
Then we \emph{``squeeze''} the profile,  
\beq
\cA_{p}^{sq}(U) = k 
\dfrac {p}{2\cosh^2 (p\, U)} \,, 
\qquad
k=\const,
 \qquad
 p \geq 1,
\label{PTp}
\eeq
where $p\geq 1$ is a real parameter,
shown  in FIG.~\ref{squeezePT}. 
Letting $p\to\infty$  an \emph{impulsive wave} is obtained, which is known to \emph{support only VM trajectories}  \cite{PodolskyVeselyCz98,Podolsky:1998sa,
Steinbauer97,Steinbauer98,Khare,P-SSS17,ZDHimpulsive,SteinbauerComm, EZHRev,CEM-SUSY}.

DM  can be restored separately for each fixed value of the deformation parameter $p$ by allowing  for $p$-dependent amplitude, 
$ k=k_p\,.$  
The price to pay is however that the new amplitudes \eqref{pDMk} diverge when $p\to\infty$ ,
 and thus, consistently with \eqref{tildearea}, \emph{no impulsive wave} is obtained.
 \goodbreak
 
\section{Squeezing the \PT profile}\label{squeezedSec}
 
We first recall some properties of impulsive waves \cite{Steinbauer97,Steinbauer98,Pod98,PodolskyVeselyCz98,Podolsky:1998sa,P-SSS17,ZDHimpulsive,SteinbauerComm}. 
While gravitational waves with smooth profile have oscillating geodesics of both DM and VM type \cite{ZelPol, DM-1,Jibril,DM-2,Sila-PLB}, impulsive waves which have a Dirac-delta profile,
 \beq
 \cA_{\infty}(U)= k\,\delta(0)\,,  \qquad k=\const. 
 \label{impprof}
 \eeq
 support only VM geodesics \cite{Steinbauer97,Steinbauer98,Pod98,PodolskyVeselyCz98,Podolsky:1998sa,P-SSS17,ZDHimpulsive,SteinbauerComm,Khare,CEM-SUSY}.
 To understand how this comes about, 
 we note that in both regions $\cI_{-}=\{U< 0\}$ and $\cI_{+}=\{U> 0\}$ the motion is free.
Requiring that the trajectory  be continuous at 
$U=0$~\footnote{Continuity, unlike smoothness, is required also for approximate solutions \cite{Benin,Approxi}.}, the \StL eqn \eqref{SLeq} with the initial conditions \eqref{initcond} then implies that the geodesics are 
 of the VM-type with amplitude-depending slope \cite{Steinbauer97,Steinbauer98,PodolskyVeselyCz98,Podolsky:1998sa,P-SSS17,ZDHimpulsive,SteinbauerComm},
\begin{equation}
X\left(U\right) =\left\{ 
\begin{array}{cl}
X_{0},&\text{ \ }U<0 \\ 
(1-kU)X_0,&\text{ \  }U>0
\end{array} \;%
\right. \,.
\label{exactImpulsive}
\end{equation}

The area below the impulsive profile $\cA_{\infty}$  is the amplitude, 
$ 
\displaystyle\int\!\!\cA_{\infty}(U)dU=k\,.
$ 
  
\begin{figure}[h]
\includegraphics[scale=.9]{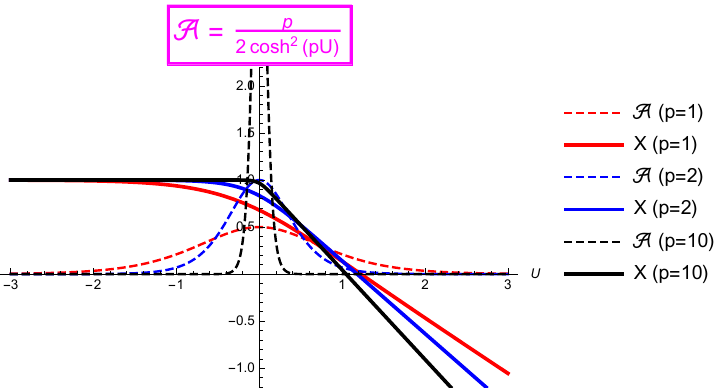}\vskip-3mm\caption{\textit{\small For the squeezed \PT profile \eqref{areak} all geodesics are of the VM type. Their slopes increase with $p$.}
\label{VMsqueeze} }
\end{figure}

Impulsive profiles can be obtained by deforming smooth ones. We can, for example,  ``squeeze'' the \PT profile
 by letting $p\to\infty$ in \eqref{PTp}.   
The squeezing parameter $p$  in the
 argument of $\cosh$ in the denominator determines the width of the bell~: large $p$ implies narrow profile, as shown in FIG.\ref{squeezePT}.  The  $p$  in the numerator implies in turn that the area below $\cA_p(U)$ remains constant for all $p$,  
\beq
\int_{-\infty}^{\infty}\!\!\cA_p(U)dU =
 k\displaystyle\int_{-\infty}^{\infty}\frac{p}{2\cosh^2 (p\,U)}dU \;= \; k=\const\,,
 \label{areak}
\eeq 
consistently with \eqref{impprof}.
In conclusion,
 {squeezeing}  \eqref{PTp} by letting  $p\to\infty$ while keeping $k =\const$  the profile converges to that of an impulsive wave with amplitude $k$, \eqref{impprof}, as shown in FIG.\ref{squeezePT}.
 Numerical solutions shown in FIG.\ref{VMsqueeze} confirm that keeping $k=1$ fixed
 yields VM geodesics (\ref{exactImpulsive}) whose slope increases with $p$.

\section{Profile for \DM}\label{DMSec}

Previous work \cite{DM-1,DM-2} indicates that for smooth profiles DM trajectories arise when the amplitude takes some magic value $k=k^{crit}$ \cite{DM-1,DM-2}.  
For  the  PT profile, \eqref{PTp} with $p=1$,  
for example, DM geodesics are obtained for
\beq
k^{crit}_{p=1}= 
2m(m+1)\,, \quad m=1,\, 2,\, \dots 
\label{PTDM}
\eeq 
shown in FIGs.\#9 and \#12 of \cite{DM-1}.
When $m$ is not a natural number, then  VM-type  trajectories are obtained \cite{ZEH-PR}.
Turning to our squeezed model \eqref{PTp} 
 with $p\geq 2$, our clue is to allow the amplitude  to become  $p$-dependent, $k=k_p$. The geodesic equation for the squeezed PT profile \eqref{PTp} is,
\begin{equation}
\dfrac {d^2 X}{dU^2} + \widetilde{\cA}_p(U) X 
 \; = \; \dfrac {d^2 X}{dU^2} + k_p \dfrac{p}{2\cosh^2(p\,U)} X  = 0\,.
 \label{pPTSL}
\end{equation}
Putting here
\beq
U \to \tilde U = p\, U\,,
\label{UpU}
\eeq
 this becomes
\begin{equation}
\dfrac {d^2 X }{d\tilde U^2} + \left(\dfrac{k_p}{p}\right) \dfrac {1}{2\cosh^2 \tilde U}\, X = 0\,,
\label{cosheq}
\end{equation}
which is of the usual PT form with redefined amplitude $k_p \to k_p/p$. For each fixed $p$ a DM solution is thus obtained when 
\begin{equation}
k^{crit} \equiv k_p = 2m(m+1) \,p\,, \quad m=1,2, \dots
\label{pDMk}
\end{equation}
which generalises \eqref{PTDM} to any $p>0$.
In conclusion, DM trajectories can be found for the \emph{enhanced \PT profile} highlighted by the $p^2$ factor in the amplitude,
\begin{equation}
\bigbox{
\widetilde{\cA_p}(U)\; = \; k_p\,\frac{p}{2\cosh^2 (p\,U)} \; = \; \dfrac{m(m+1)}{\cosh^2 (p\,U)}\,p^2\, 
. }
\label{p2DMk}
\end{equation}

\begin{figure}[h]
\includegraphics[scale=.47]{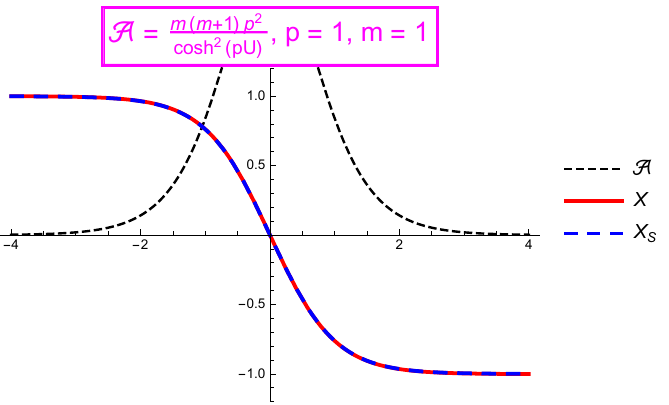}\;
\includegraphics[scale=.47]{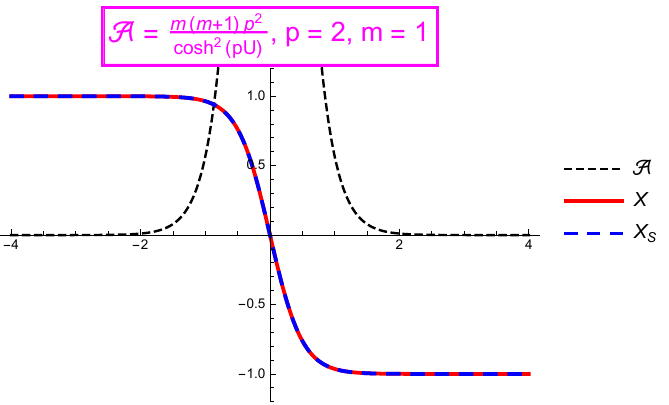}\;
 \includegraphics[scale=.47]{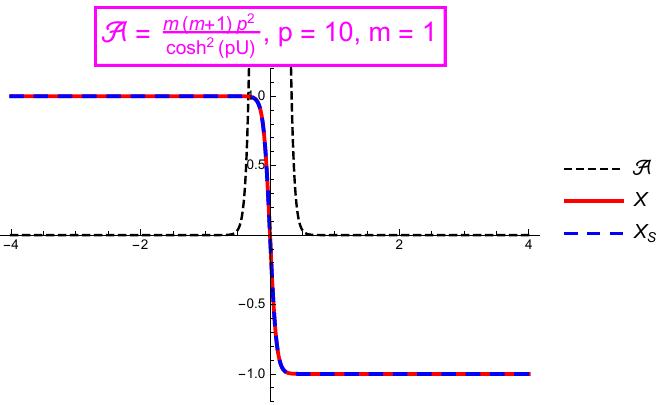}

\vskip-3mm\caption{\textit{\small DM geodesics for the
enhanced \PT profile \eqref{p2DMk} with {\bf odd} wave number $m=2\ell+1$ shown here for $m=1$. Increasing the deformation parameter the trajectories remain of the DM type, only become steeper and letting $p\to \infty$
converge to an antisymmetric step function. The \red{\bf red} curve is the {numerical} solution, and dashed \blue{\bf blue} line is the analytical one.}
}
\label{psqueeze-m1} 
\end{figure}

\begin{figure}[h]\hskip-2mm
\includegraphics[scale=.48]{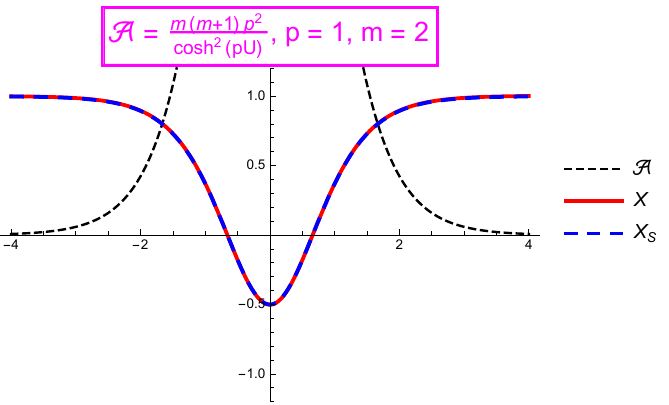}
\includegraphics[scale=.48]{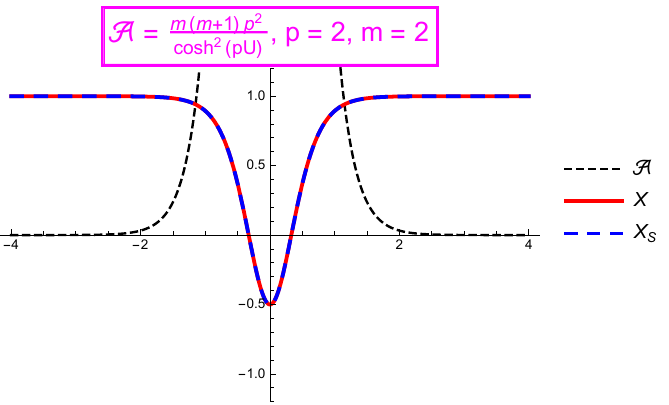}
 \includegraphics[scale=.48]{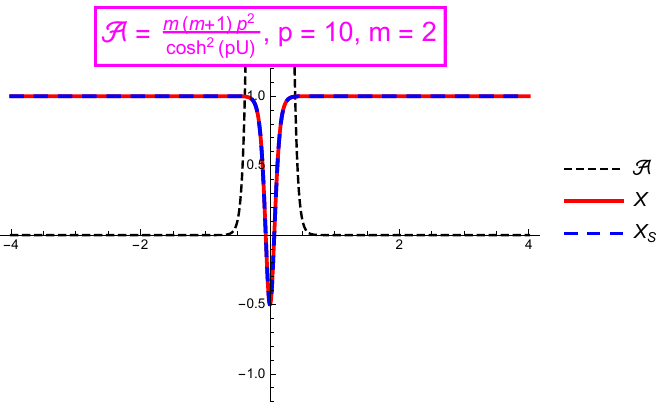}
\vskip-6mm\caption{\textit{\small For {\bf even} wave number $m=2\ell$ shown here for $m=2$.
The geodesics of the squeezed PT profile  
are ``DM with no final displacement''.  The \red{\bf red} curve is the {numerical} solution, and dashed \blue{\bf blue} line is the \blue{analytical} one.} %
\label{psqueeze-m2}
} 
\end{figure}

The trajectories, shown in FIGS.\ref{psqueeze-m1}-\ref{psqueeze-m2}, are given by {``squeezed''} Legendre polynomials \cite{DM-1}. Their behaviour is consistent with their parity-dependent odd or even symmetry  
 \beq
 \tilde{X}(-\infty)=\tilde{X}_0 \;\to\; \tilde{X}(+\infty)=
 (-1)^m\tilde{X}_0\,,
 \label{PTmjump}
 \eeq
where the  label $p$ was dropped.
The analytic results are confirmed numerically. 

\goodbreak

\medskip
\begin{figure}[h]
\includegraphics[scale=.47]{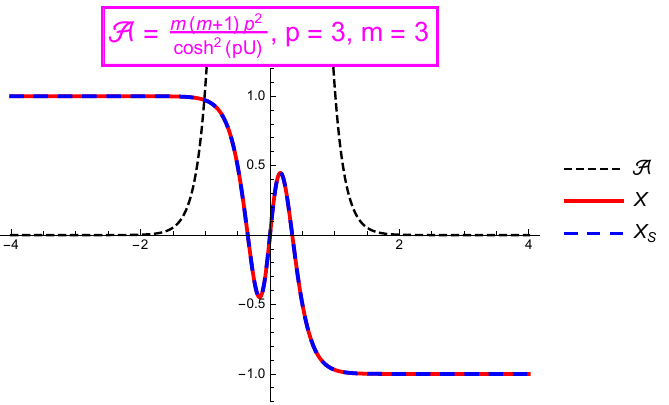}\,
\includegraphics[scale=.47]{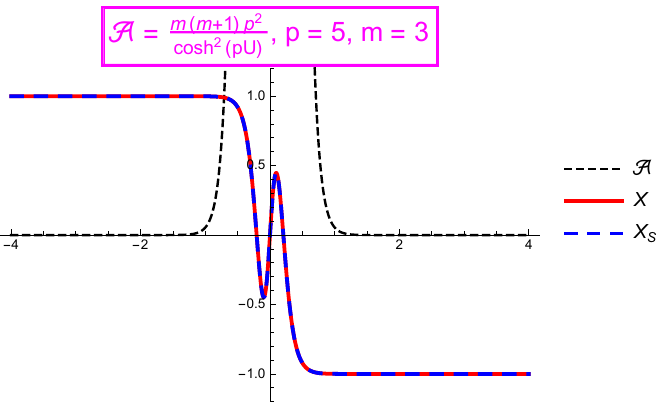}\,
\includegraphics[scale=.47]{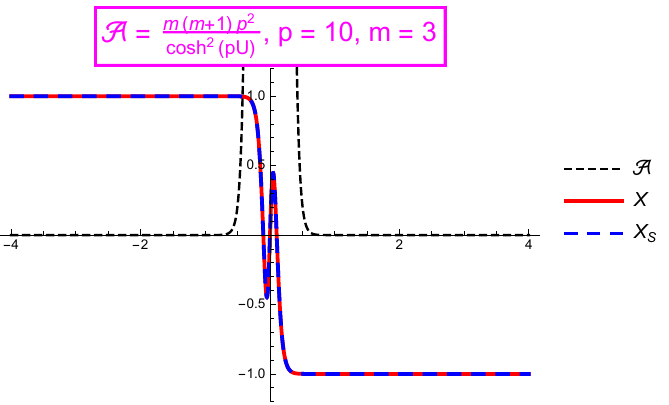}
\vskip-5mm\caption{\textit{\small  
DM trajectories for higher wave number and  deformation parameters $m$ and $p$ for the enhanced model \eqref{p2DMk}. 
} %
\label{sq-m2m3p3}
} 
\end{figure}

Similar behaviour is found for higher wave numbers and deformation parameters, see FIG.\ref{sq-m2m3p3}.
We emphasise, though, that the enhanced DM profile \eqref{p2DMk} does \emph{not}
converge to that of an impulsive wave~: the area below the profile  diverges due to the extra $p$ factor, 
\beq
\int_{-\infty}^{\infty}\!\widetilde{A}_p\,dU =m(m+1)\,p \to \infty\,,
\label{tildearea}
\eeq
violing the impulsive conditions \eqref{impprof} or  \eqref{areak}. The enhanced model \emph{is} DM but is \emph{not} impulsive.

\section{A different profile with identical impulsive limit}\label{Anotherprof}

\begin{figure}[h]
\includegraphics[scale=.73]{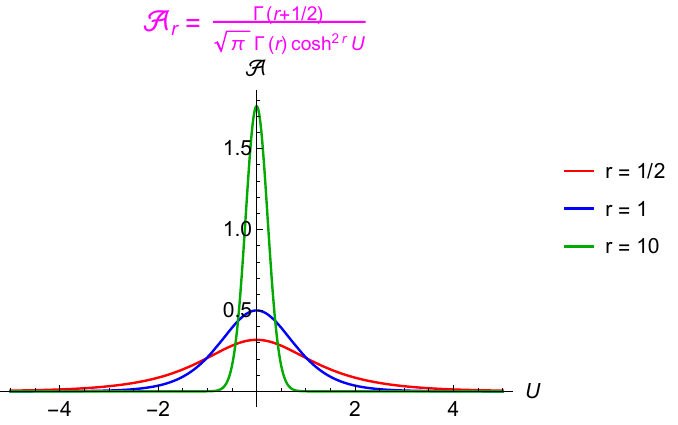}
\vskip-5mm\caption{\textit{\small 
Fixing the  amplitude at $k=1$ and letting $r\to \infty$ the profile \eqref{coshrProf} approaches that of a \PT-based impulsive wave, \eqref{impprof} in FIG.\ref{squeezePT}. 
}
\label{coshrU} }
\end{figure}

A remarkable result in \cite{PodolskyVeselyCz98,Pod98,Steinbauer97}
says that the impulsive limit is  independent of the initial profile  as it is nicely illustrated by comparing the results discussed above with what happens for another deformation
of PT, 
\beq
\widehat{\cA}_{r}(U) = 
k_r\delta_r(U)
= k_r\frac{1}{\sqrt{\pi}}\,\frac{\Gamma\!\left(r+\half
\right)}{\Gamma\!\left(r\right)}
\frac{1}{\cosh^{2r}U}\,,
\qquad 
 r = 1\,, \, \dots 
\label{coshrProf}
\eeq
shown in FIG.\ref{coshrU}  for $k_r=1$.
The $r=1$  profile is \eqref{PTGprof}.
%
The main results, whose details will be presented in \cite{ZEH-PR} says that~:
\benu
\item
The regularised kernels
\beq
\delta_r(U)
= \frac{1}{\sqrt{\pi}}\,\frac{\Gamma\!\left(r+\half
\right)}{\Gamma\!\left(r
\right)}
\frac{1}{\cosh^{2r}U}
\qquad\text{\small verify}\qquad
\lim_{r\to\infty}\delta_r(U)=\delta(U)\,
\label{PM-2}
\eeq
in the sense of tempered distributions. 
For $r=1$ we have $ \Gamma (3/2) / \Gamma(1) = \sqrt{\pi}/2$ and we recover the initial \PT solutions in \cite{DM-1}. The area below the profile can be calculated  analytically \cite{AbramowitzStegun}. The result is consistent with \eqref{impprof}.

\item
Developing 
$
{\Gamma(r+1/2)}/{\Gamma(r)}  
$ 
for large $r$,  we get approximately
\beq
\delta_r(U)\;\sim\;
\frac{1}{\sqrt{\pi}}\frac{1}{\cosh^{2r}U}\;
\sqrt{r}\left(1 -\frac {1}{8r} + \dots\right)\,.
\eeq
Thus the profiles decays exponentially.

\item
Near the origin, we recover the  Gaussian  profile
\beq
\delta_r(U)\approx\sqrt{\frac{r}{\pi}}\exp\left(-{r}\,U^2\right).
\eeq

\begin{figure}[h] 
\includegraphics[scale=.18]{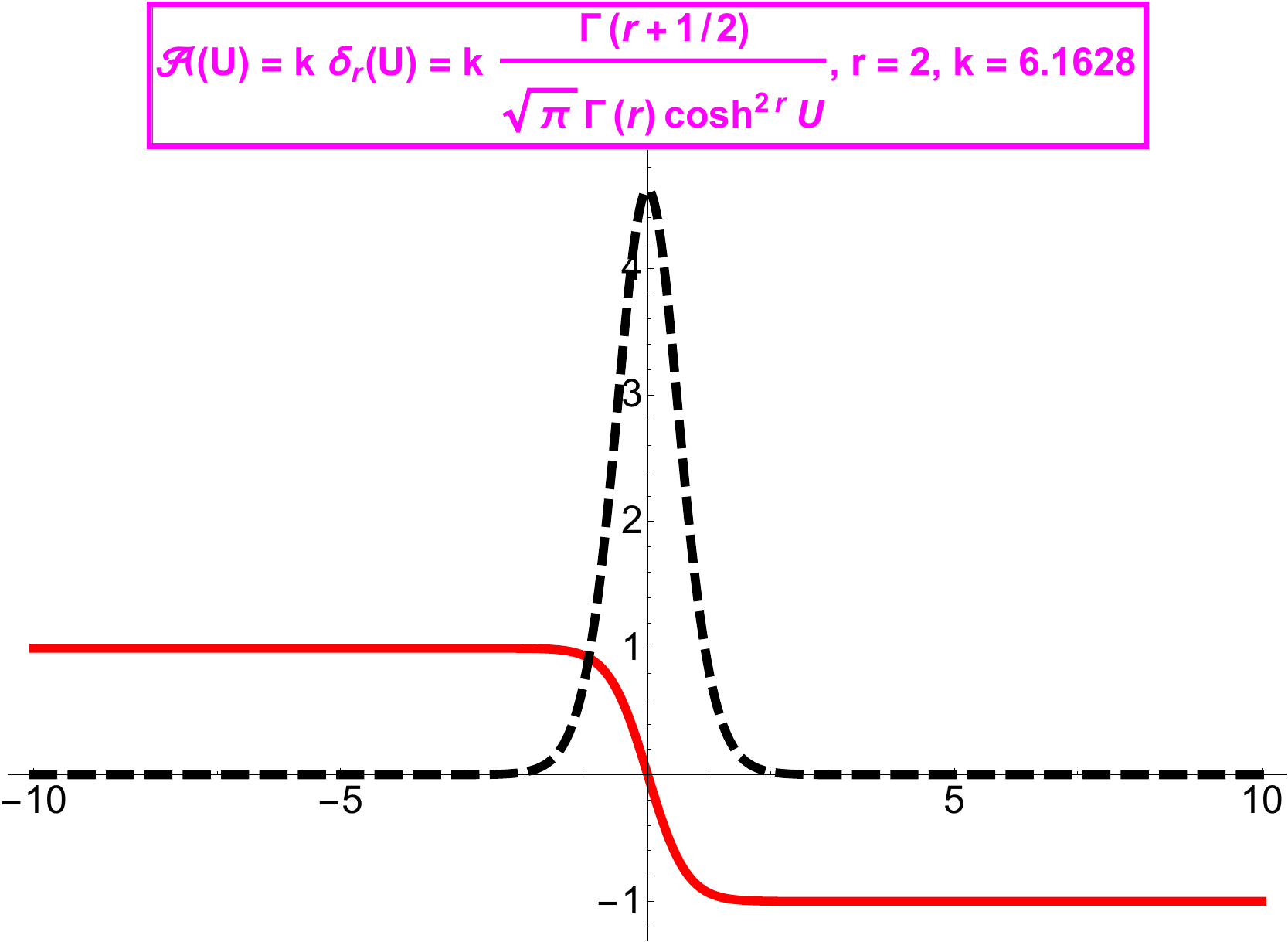}
\includegraphics[scale=.18]{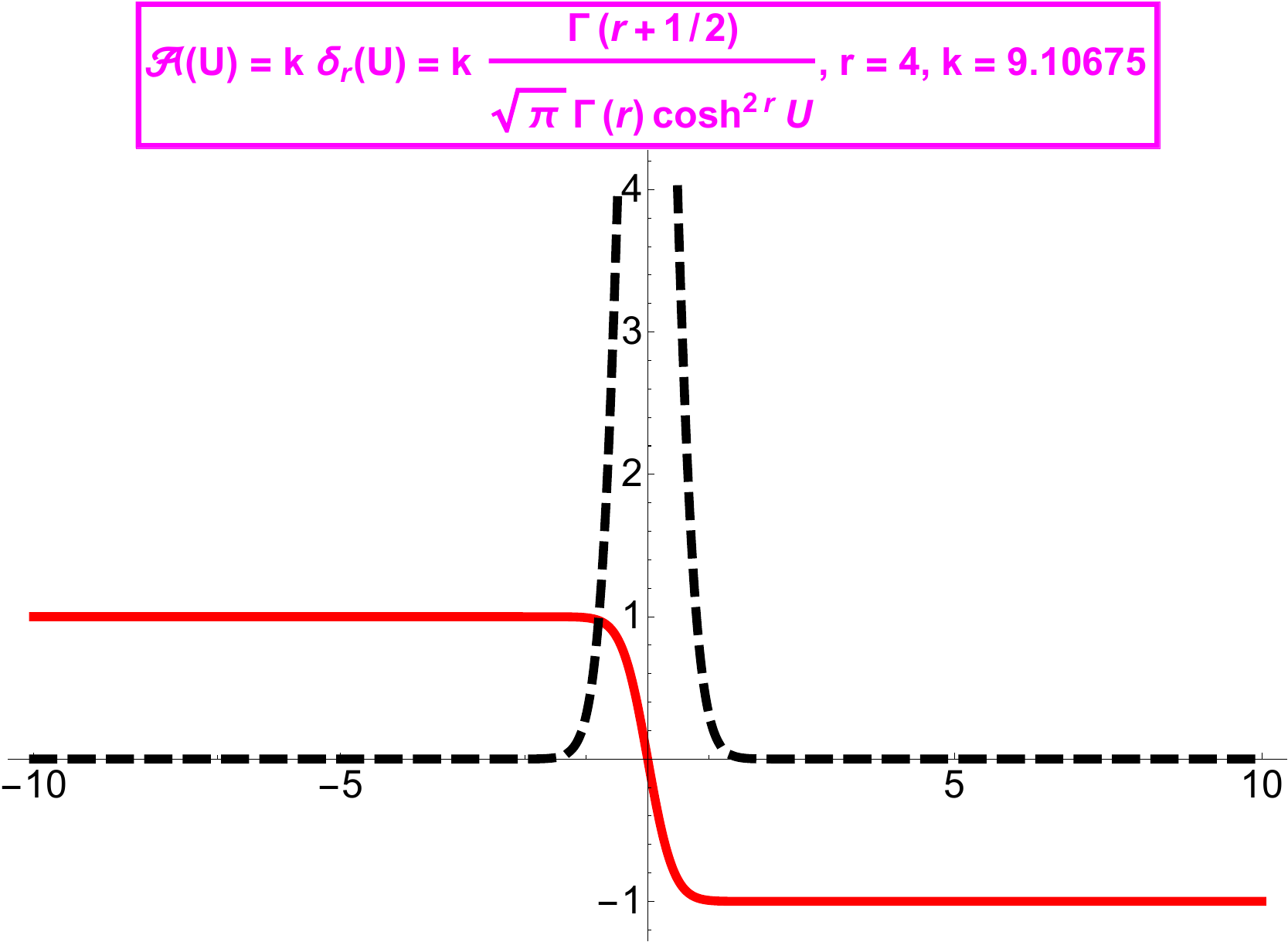}
\includegraphics[scale=.18]{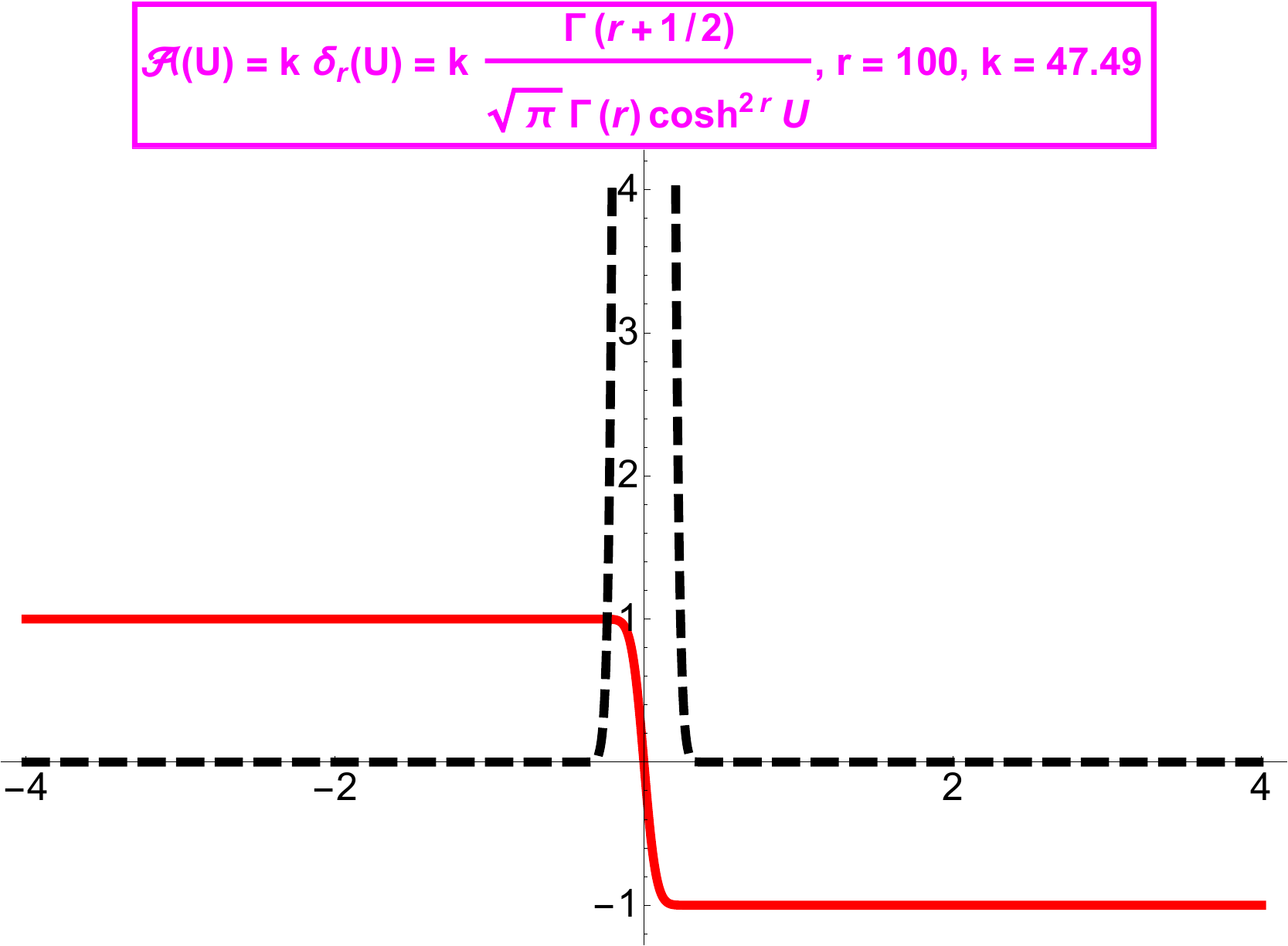}
\vskip-3mm\caption{\textit{\small The DM trajectories for the profile \eqref{coshrProf} become steeper when the deformation parameter $r\to\infty$. To be compared with enhanced PT in FIG.\ref{psqueeze-m1}. 
}
\label{OtherDM} }
\end{figure}

\item
The geodesic equations can be solved  numerically by following our strategy proposed in \cite{Approxi}.
DM behaviour is recovered by trading the constant amplitude $k$ for ``magic'' parameters \cite{Approxi}. 
For DM with $m=1$ wave number, we found, 
\beq\hskip-4mm
\hat{k}_{2} = 6.1628\,,\;\;
\hat{k}_{3} = 7.77185\,,\;\;
\hat{k}_{10} = 14.7834\,,\;\;
\hat{k}_{20} = 21.0906 ,\,\dots 
\label{OtherDMk}
\eeq
illustrated in FIG.\ref{OtherDM}, to be compared  with the PT values \eqref{pDMk}. For $\hat{k}_1 = 4$ the \PT trajectory \eqref{psqueeze-m1} is recovered.
\goodbreak
\eenu
Thus the area integral 
\beq
\widehat{\rm Area\,}_{r}=
\int_{-\infty}^{\infty} \hat{k}_r\delta_r
= k_r \frac{1}{\sqrt{\pi}}\,\frac{\Gamma\!\left(r+\half
\right)}{\Gamma\!\left(r
\right)}
\frac{1}{\cosh^{2r}U} dU
= \hat k_r
\label{OtherArea} 
\eeq
For $m=1$,
$\hat{k}_r \approx 4.7 \sqrt {r}$ which diverges confirming the impulsive character. Comparison with  \eqref{tildearea} shows that ${r}$ plays here the r\^ole of $p^2$ in the PT model.

\section{$UV$ scaling and chronoprojective structures}
\label{chronoSec}

The symmetries of \eqref{Bplanewave} are the isometries  which leave both the metric \emph{and} $\xi$ invariant,
\beq
L_Yg_{\mu\nu} = 0\, \aand L_Y\xi= 0\,.
\label{Bisom}
\eeq
For a free particle, $\cA=0$, it is spanned by the usual Galilei generators centrally extended by $\p/\p{V}$,
called the Bargmann algebra \cite{DBKP,DGH91}.
For the harmonic profile, which is our particular interest here, the isometries belong to the Carroll algebra with no rotations \cite{Sou73,Carroll4GW}. 

The time dilation \eqref{UpU} has an unusual behavior:  it lifts to Bargmann space as an isometry, however it does \emph{not} belong to the Bargmann group. It belongs instead to its $1$-parameter extension called the \emph{Chronoprojective group} 
 \cite{DuvalThese,Burdet82,BDP-LMP,PBDChrono}, obtained by generalizing \eqref{Bisom}
 \cite{ZCEH-chrono}. Chrono-projective transformations are
 isometries which preserve only the direction of $\xi$,
\beq
L_Yg_{\mu\nu} = 0\, \aand L_Y\xi = \psi \xi  \,.
\label{UVChrono}
\eeq
For a full account the reader is referred to the literature.

Our particular interest here concerns the $1D$ (E-D) oscillator metric \eqref{Bplanewave}  with  $\widetilde{\cA}$ given in \eqref{pPTSL}.
The $U$-rescaling \eqref{UpU} lifts to the Bargmann space,
\begin{equation} 
X \to X,\quad
U \to p\, U, \quad  V \to p^{-1} V,
\quad\text{\small generated by}\quad
Y = U \partial_U - V \partial_V\,.
 \label{UVboost}
\end{equation}
It is  \emph{not} a symmetry in general. 
For $\psi\neq0$  such isometries do {not} belong to the Bargmann group. For an arbitrary profile $\cA\neq 0$,
only the $uu$-component,
$ 
\mathcal{L}_Y g_{uu}=(U\partial_U+2)\,\cA X^2\,, 
$ 
is effected,
the others vanish identically. For our harmonic profile, \eqref{UVChrono} is satisfied with $\psi=1$. 
Some examples~:
\begin{enumerate}
\item On a harmonic oscillator, $\cA(U)=\const$, the typical solution $X_{p=1}(U) = \cos(U)$ is taken into $X_{p}(U) = \cos(p\, U)$ by and UV boost. Its lift is  a geodesic for the metric 
\begin{equation}
\label{tildeg} 
\tilde{g}_p =  dX^2 + 2 dU dV +p^2 \cA(p\, U) X^2 dU^2. 
\end{equation}

\item For the PT profile \eqref{PTGprof},  DM solutions $X_{p=1}(U)$ are taken to $X(p\, U)$. The  trajectory is distorted however remains a DM geodesic, as we have seen in sect.\ref{DMSec}.

\item
 UV-boosts are symmetries  for the singular potential 
\beq
\cA(U)=\frac{a}{U^2}\,,
\quad
a=\const\,.
\label{inverseU2}
\eeq
 \cite{AndrzejewskiP-1,AndrzejewskiP-2,Ilderton,ZCEH-chrono,Benin,ZhaoCao}~:  
the profile is plainly invariant under the time rescaling $U \to p \,U$.
 
\end{enumerate}
\goodbreak

\section{Conclusion}\label{Concl}

For general parameters,  
 gravitational waves  exhibit the \VM \cite{Ehlers,BraTho,Favata}. However for certain ``magic'' values of the amplitude they may support the \DM \cite{ZelPol}.    
In this paper  we revisit their relation in a new arena. 

The key   
is the interplay between three $p$ factors in sects.\ref{squeezedSec} and \ref{DMSec}~: the first $p$, 
 \eqref{UpU} 
in the argument of  $\cosh$ in \eqref{cosheq}, 
  squeezes the profile to the origin without changing the amplitude. The second $p$ in the numerator in \eqref{pPTSL} 
  compensates the loss of area by increasing the amplitude,  \eqref{areak}, 
 as  required for impulsive waves, \eqref{impprof}. 
   But then we get only \VM\!, as shown in FIG.\ref{VMsqueeze}. DM can be restored for fixed $p$ by further increasing the amplitude as in \eqref{pDMk},
 yielding  \eqref{p2DMk}. 
   But then the area diverges, \eqref{tildearea},  
 and the wave is not more impulsive.
 
While completing our research, we  came across a remarkable series of early papers highlighted by
\cite{Pod98,PodolskyVeselyCz98,Steinbauer97}.
Below we summarise shortly their results.
  Podolsky, in his unjustly ignored  paper \cite{Pod98},
   starts with the \PT- type  profile $\cA=\frac{a}{\cosh^2(b\, U)}$, where $a$ and $b$ are constants, he calls a ``smooth asymptotic sandwich wave" \cite{Pod98,PodolskyVeselyCz98}. 
Then he finds the geodesics in terms of hypergeometric functions, which, for large $U$, become linear, consistently with \cite{DM-1}. 
For special values  $k_{m}=2m(m+1)$
of the \PT profile he finds their exact form,  consistent with \cite{DM-1}.
 At last, he notices that taking the large-$U$ limit, the profile becomes impulsive,
consistently with our sec.\ref{squeezedSec}~\footnote{DM trajectories are in fact plotted in FIG.1 of \cite{Pod98}. 
 Their repulsive components diverge, though, and are therefore discarded. In \cite{DM-1} they are called ``half DM".}.

 The models we study here have interesting applications in the Newtonian version of Friedmann-Lema\^\i tre cosmology \cite{DuvalThese,BDP-LMP,
 PBDChrono, FST-Astro}.  In E-D terms, it corresponds to profile
 \beq
 \cA^{FL} \propto 
  \big(\frac{\ddot{\beta}}{\beta}\big)\,,
 \label{FLprofile}
 \eeq 
 where the scale function $\beta$  is related to the Hubble parameter, $H=\dot{\beta}/{\beta}$.
A remarkable aspect of the cosmological application is that the initially attractive oscillator potential has an inflexion point after which it becomes repulsive, consistent with expansion of the Universe \cite{BDP-LMP,PBDChrono,FST-Astro}.

Further related aspects including SUSY  are studied in
 \cite{Khare,CEM-SUSY,ZEH-PR}.

\begin{acknowledgments} 
PMZ was partially supported by the National Natural Science Foundation of China (Grant No. 12375084).
ME is supported by The Scientific and Technological Research Council of Turkey (T\"UBITAK) under grant number 125F021. PAH and PMZ thank Bilgi U. for hospitality and  T\"UBITAK for financial support. Discussion on related cosmological models is acknowledged to R. Triay.
\end{acknowledgments}
\goodbreak

\end{document}